\documentclass[11pt,preprint]{aastex}

\newcommand{\ebr}{E$_{{\mbox{br}}}$}
\newcommand{\spe}{S$_{{\mbox{e}}}$}
\newcommand{\spl}{S$_{{\mbox{l}}}$}

\shorttitle{Cluster Mass and Mass-to-Light Profiles}
\shortauthors{Katgert, Biviano \& Mazure}

\received{2003 July 3}
\begin{document}


\title{The ESO Nearby Abell Cluster Survey\footnote{Based on 
observations collected at the European Southern Observatory (La Silla, 
Chile) \protect\\ http://www.astrsp-mrs.fr/www/enacs.html} \protect\\ 
XII. The mass and mass-to-light-ratio profiles of rich clusters}


\author{Peter Katgert}
\affil{Sterrewacht Leiden, Postbus 9513, Niels Bohrweg 2, 2300 RA Leiden, 
The Netherlands}
\email{katgert@strw.leidenuniv.nl}
\author{Andrea Biviano}
\affil{INAF/Osservatorio Astronomico di Trieste, 
Via Tiepolo 11, I-34131 Trieste, Italy}
\email{biviano@ts.astro.it}
\and 
\author{Alain Mazure}
\affil{OAMP/Laboratoire Astrophysique de Marseille, Traverse du Siphon, 
Les Trois Lucs, BP 8, F-13376 Marseille Cedex, France}
\email{alain.mazure@oamp.fr}

\begin{abstract}
We determine the mass profile of an ensemble cluster built from 3056
galaxies in 59 nearby clusters observed in the ESO Nearby Abell
Cluster Survey. The mass profile is derived from the distribution and
kinematics of the Early-type (elliptical and S0) galaxies only, with
projected distances from the centers of their clusters \mbox{$\leq 1.5
\, r_{200}$}. These galaxies are most likely to meet the conditions
for the application of the Jeans equation, since they are the oldest
cluster population, and are thus quite likely to be in dynamical
equilibrium with the cluster potential. In addition, the assumption
that the Early-type galaxies have isotropic orbits is supported by the
shape of their velocity distribution. For galaxies of other types (the
brightest ellipticals with $M_R \leq -22+5 \log h$, and the early and
late spirals) these assumptions are much less likely to be
satisfied. For the determination of the mass profile we also exclude
Early-type galaxies in subclusters. Application of the Jeans equation
yields a {\em non-parametric estimate} of the cumulative mass profile
$M(<r)$, which has a logarithmic slope of $-2.4 \pm 0.4$ in the
density profile at $r_{200}$ (approximately the virial radius). We
compare our result with several analytical models from the literature,
and we estimate their best-fit parameters from a comparison of
observed and predicted velocity-dispersion profiles. We obtain
acceptable solutions for all models (NFW, Moore et al. 1999, softened
isothermal sphere, and Burkert 1995). Our data do not provide
compelling evidence for the existence of a core; as a matter of fact,
the best-fitting core models have core-radii well below
100~$h^{-1}$~kpc. The upper limit we put on the size of the
core-radius provides a constraint for the scattering cross-section of
dark matter particles. The total-mass density appears to be traced
remarkably well by the luminosity density of the Early-type
galaxies. On the contrary, the luminosity density of the brightest
ellipticals increases faster towards the center than the mass density,
while the luminosity density profiles of the early and late spirals
are somewhat flatter than the mass density profile.
\end{abstract}


\section{Introduction}
\label{s-intro}
Many attempts have been made to determine the amount and distribution
of dark matter in clusters, since Zwicky (1933, 1937) and Smith (1936)
concluded that the mass implied by the sum of the luminosities of the
galaxies falls short of the total mass by as much as a factor of 10.
In recent years, cosmological simulations have shown that dark matter
halos have a universal density profile (Navarro, Frenk, \& White 1996,
1997; NFW hereafter), but its precise form is still debated (Moore et
al. 1999, M99 hereafter), while the 'universality' of the mass density
profile has also been questioned (Jing \& Suto 2000; Thomas et
al. 2001; Ricotti 2002). The issue is critical, since knowledge of the
total mass (visible and dark, baryonic and non-baryonic) of clusters,
and of its distribution (also in relation to the light distribution),
gives important clues about the formation process of the clusters
(e.g. Crone, Evrard, \& Richstone 1994; Jing et al. 1995), and of the
galaxies in them (e.g. Mamon 2000), as well as on the nature of dark
matter (see, e.g., Natarajan et al. 2002a).

The dark matter can be `weighed' and `imaged' in three ways. First,
the gravitational lensing of distant objects yields an estimate of the
projected mass distribution in the cluster (see e.g. Tyson, Wenk, \&
Valdes 1990; Squires et al. 1996). The mass density profile can
subsequently be derived assuming the geometry of the cluster. Lensing
mostly works for clusters at intermediate and large distances since
the lensing equation is unfavourable for nearby clusters, and only few
nearby cluster lenses are known (see e.g. Campusano, Kneib, \& Hardy
1998). Some results from gravitational lensing observations are
consistent with the NFW profile (e.g. Athreya et al. 2002; Clowe \&
Schneider 2001), while others favour flatter mass density profiles,
such as the isothermal sphere (see Fischer \& Tyson 1997; Sand, Treu,
\& Ellis 2002; Shapiro \& Iliev 2000; Taylor et al. 1998).

A second method uses the distribution of the hot, X-ray emitting, gas
in the cluster potential, and the radial variation of the temperature
of the gas (see e.g. Hughes 1989; Ettori, De Grandi, \& Molendi
2002). Recently, temperature maps have become available for a sizable
number of clusters (see e.g. Markevitch et al. 1998; Irwin \& Bregman
2000; De Grandi \& Molendi 2002). However, the uncertainties in the
X-ray temperature profiles are still substantial (see e.g. Irwin,
Bregman, \& Evrard 1999; Kikuchi et al. 1999), with corresponding
uncertainties in the total mass and the mass profile. Moreover, X-ray
observations in general sample only the inner cluster regions, even
with the new class of X-ray satellites (see, e.g., Pratt \& Arnaud
2002). Cooling can also play a r\^ole, although it is probably not
dominant (Suginohara \& Ostriker 1998). Most recent results from X-ray
observations indicate consistency with the NFW or the M99 profile
(e.g. Allen, Ettori, \& Fabian 2001; Allen, Schmidt, \& Fabian 2002;
Markevitch et al. 1999; Pratt \& Arnaud 2002; Tamura et al. 2000), but
a flatter mass density profile like that of the isothermal sphere
seems to be preferred in some clusters (Arieli \& Rephaeli 2003;
Ettori et al. 2002). Generally, the distribution of the the X-ray
emitting gas is found to be flatter than that of the total cluster
mass (e.g. Allen et al. 2001, 2002; Markevitch \& Vikhlinin 1997;
Nevalainen, Markevitch, \& Forman 1999, Pratt \& Arnaud 2002).

The third, and most traditional way to probe the dark matter content
and its distribution in clusters is through the kinematics and spatial
distribution of 'tracer particles' moving in the cluster potential.
The virial theorem applied to the galaxy population gives an estimate
of the total mass but, since the data in general comprise only the
central part of a cluster, projection effects and possible anisotropy
of the velocity distribution must be taken into account for an
accurate estimate (see e.g. Heisler, Tremaine, \& Bahcall 1985; The \&
White 1986). In addition, the virial theorem assumes that mass follows
light and if this assuption is not justified that can have a large
effect on a cluster mass estimate (see, e.g., Merritt 1987).

Recently, Diaferio \& Geller (1997; see also Diaferio 1999) used the
caustics in the plane of line-of-sight velocities vs. clustercentric
distances to estimate the amplitude of the velocity field in the
infall region. This yields an estimate of the escape velocity at
several distances, and thus of the mass distribution. This method has
the advantage of being free of the assumption of dynamical
equilibrium, and hence can be (and has been) used to constrain the
cluster mass distribution well beyond the virialization region
(Geller, Diaferio, \& Kurtz 1999; Reisenegger et al. 2000; Rines et
al. 2000, 2001, 2002; Biviano \& Girardi 2003). However, systematic
uncertainties limit the accuracy in the mass determination to $\sim
50$\% (Diaferio 1999). Results obtained with this method indicate
consistency with the NFW or M99 profiles, or strongly constrain or
rule out the isothermal sphere model.

The most detailed form of the 'kinematical weighing' uses the
distribution and kinematics of the galaxies to estimate the mass
distribution, through application of the full Jeans equation of
stellar dynamics (see, e.g., Binney \& Tremaine 1987, BT hereafter).
This method requires knowledge of the orbital characteristics of the
galaxies. A first analysis along these lines was made for the Coma
cluster by Kent \& Gunn (1982). Merritt (1987) used the same data to
estimate the orbital anisotropy of the galaxies for various
assumptions about the dependence of the mass-to-light ratio ${\cal
M}/{\cal L}$ on radius.

The shape of the mass profile of a cluster would follow directly from
the projected luminosity density profile of the galaxies if the
mass-to-light ratio ${\cal M}/{\cal L}$ were constant. Evidence in
favour of a constant ${\cal M}/{\cal L}$ comes from the analyses of
Carlberg, Yee, \& Ellingson (1997a), van der Marel et al. (2000), and
Rines et al.  (2001). Carlberg et al.  (2001) instead find an
increasing ${\cal M}/{\cal L}$-profile in groups from the CNOC2
survey, while both Rines et al. (2000) and Biviano \& Girardi (2003)
argue for a decreasing ${\cal M}/{\cal L}$-profile.

The shape of the cluster ${\cal M}/{\cal L}$-profile seems to depend
on which class of cluster galaxies is used to measure the light
profile. This is hardly surprising, since the various types of
galaxies are known to have different projected distributions. These
differences are the result of the morphology-density relation, first
described by Oemler (1974) and Melnick \& Sargent (1977), and
described more fully by Dressler (1980) and, lately, Thomas \& Katgert
(2003, paper X of this series). In clusters, this relation produces
differences in the radial distribution of the various galaxy classes,
and those are related to differences in the kinematics. Evidence for
the relation between spatial distribution and kinematics for different
cluster galaxy populations was e.g. found in the Coma cluster (Colless
\& Dunn 1996), in CNOC clusters at $z \approx 0.3$ by Carlberg et
al. (1997b), in clusters observed in the ESO Nearby Abell Cluster
Survey (ENACS, hereafter; de Theije \& Katgert 1999, paper VI; Biviano
et al. 2002, paper XI), as well as in other clusters (see, e.g., Mohr
et al. 1996; Adami, Biviano, \& Mazure 1998a). Galaxies with emission
lines (ELG) provide an extreme example of the effect. The ELG are less
centrally concentrated and have a higher r.m.s line-of-sight velocity
than the galaxies without emission lines. This was clearly
demonstrated by Biviano et al. (1997, paper III), using 75 ENACS
clusters.

To study the mass profile in detail, one must combine the data for
many galaxy systems, as was done by e.g., Biviano et al. (1992),
Carlberg et al. (1997a, 1997b, 1997c, 2001), van der Marel et
al. (2000), Biviano \& Girardi (2003), and in papers III, VI, VII
(Adami et al. 1998b), IX (Thomas, Hartendorp, \& Katgert 2003), X, and
XI of this series. For an `ensemble' cluster, built from 14 clusters
with redshifts between about 0.2 and 0.5, with a total of 1150 CNOC
redshifts, Carlberg et al. (1997c) found that the number density
profile is consistent with the NFW mass density profile. This result
was confirmed with a more detailed analysis by van der Marel et al.
(2000). Using the ENACS dataset, in paper VII we found that the number
density profile of a composite cluster of 29 nearby ACO clusters with
smooth projected galaxy distributions did not show the central cusp of
the NFW mass profile, in particular when the brightest galaxies are
removed from the sample. This could mean that ${\cal M}/{\cal L}$
increases towards the center. Recently, Biviano \& Girardi (2003)
analysed an 'ensemble' cluster, built from 43 nearby clusters observed
in the 2dF Galaxy Redshift Survey (Colless et al. 2001). From that
sample, which has only three clusters (A957, A978 and A2734) in common
with the present sample, they concluded that both cuspy profiles of
the NFW and M99 form, and profiles with a core are acceptable, as long
as the core radius is sufficiently small.

The above results indicate the importance of a study of the mass
profile and of the radial dependence of ${\cal M}/{\cal L}$. In this
paper we present such a study, based on a sample of rich nearby
clusters observed in the ENACS. This paper is organised as follows. In
\S~\ref{s-data} we summarize the data. In \S~\ref{s-massearly} we
justify our choice to use the Early-type galaxies that are not in
substructures as tracers of the potential. In \S~\ref{s-direct} we
present the number-density and velocity-dispersion profiles of the
Early-type galaxies and we describe how we obtained a non-parametric
estimate of the cluster mass profile via direct solution of the Jeans
equation.In \S~\ref{s-model} we compare our result with models, and we
derive the best-fit models from a comparison of the observed and
predicted velocity-dispersion profiles. In \S~\ref{s-mlprof} we derive
the radial dependence of the ${\cal M}/{\cal L}$-ratio, for all
galaxies together and for the Early-type galaxies. In \S~\ref{s-disc}
we discuss our results, which are summarized in \S~\ref{s-summ}, where
we also give our conclusions. In Appendix~\ref{a-interp} we describe
our method of interloper rejection. In Appendix~\ref{a-prof} we detail
the methods by which we determined the number-density,
luminosity-density, and velocity-dispersion profiles. In
Appendix~\ref{a-assum} we review and discuss the basic assumptions
made in the determination of the mass profile.

\section{The data}
\label{s-data}

Our determination of the mass profile of rich clusters is based on
data obtained in the context of the ENACS.  The multi-object fiber
spectroscopy with the 3.6-m telescope at La Silla is described in
Katgert et al. (1996, 1998, papers I and V of this series,
respectively). In those papers, the photometry of the 5634 galaxies in
107 rich, nearby ($ z\la 0.1$) Abell clusters is also discussed. After
the spectroscopic survey was done, CCD images were obtained with the
Dutch 92-cm telescope at La Silla for 2295 ENACS galaxies. These have
yielded morphological types (Thomas 2003, paper VIII), which were used
to refine and recalibrate the galaxy classification based on the ENACS
spectra, as carried out previously in paper VI. The CCD images also
yielded structural parameters, through a decomposition of the
brightness profiles into bulge and disk contributions (paper IX).

The ENACS morphological types were supplemented with morphological
types from the literature, and subsequently combined with the spectral
types into a single classification scheme. This has provided galaxy
types for 4884 ENACS galaxies, of which 56\% are morphological, 35\%
are spectroscopic, 6\% are a combination of morphological an spectral
types, and the remaining 3\% were special in that they had an early
morphological type (E or S0) but showed emission lines in the
spectrum. These galaxy types were used to study the morphology-radius
and morphology-density relations (paper X). They also form the basis
of the study of morphology and luminosity segregation which uses
velocities as well as positions (paper XI).

In combining the data for an ensemble of many clusters, all projected
clustercentric distances were expressed in terms of $r_{200}$, as
derived from the global velocity dispersion (see e.g.  Carlberg et
al. 1997c). This ensures that we avoid, as much as possible, mixing
inner virialized cluster regions with external non-virialized cluster
regions. Similarly, all galaxy velocities, relative to the mean
velocity of their parent cluster, were expressed in terms of the
global velocity dispersion of their parent cluster, $\sigma_p$.

The ensemble cluster effectively represents each of our clusters if
these form a homologous set, and if we adopted the correct scaling.
An indication that clusters form a homologous set comes from the
existence of a fundamental plane that relates some of the cluster
global properties (Schaeffer et al.  1993; Adami et al. 1998c, paper
IV of this series). As shown by Beisbart, Valdarnini, \& Buchert
(2001) clusters with substructure tend to deviate from the fundamental
plane, thereby violating homology. The exclusion of all clusters with
even a minor amount of substructure would have greatly reduced the
size of the ensemble cluster. Instead, by eliminating the galaxies
that in their respective clusters are in substructures, we have tried
to reduce the effects of substructure in our analysis as much as
possible.

In paper XI we combined the data for 59 clusters with $z < 0.1$, each
with at least 20 member galaxies with ENACS redshifts, and with galaxy
types for at least 80\% of the members (see Table~A.1 in paper XI).
The ensemble cluster contains 3056 member galaxies, for 2948 (or 96\%)
of which a galaxy type is known. The rejection of interlopers was
based on the method of den Hartog \& Katgert (1996), which is
summarized in Appendix~\ref{a-interp}. In this ensemble cluster it was
found that galaxies in substructure have different phase-space
distributions from those outside substructure, and that this is true
for all galaxy types.

%\clearpage

\begin{table}
\caption{The numbers of galaxies outside substructure}
%\begin{flushleft}
%\begin{tabular}{|c|c|c|c|}
\begin{tabular}{cccc}
\tableline
\tableline
\ebr & Early  & \spe & \spl \\
\cline{2-4}
%$M_R \leq -22+5 \log h$ & \multicolumn{3}{c|}{$M_R > -22+5 \log h$} \\
$M_R \leq -22+5 \log h$ & \multicolumn{3}{c}{$M_R > -22+5 \log h$} \\
\tableline
34 &  1129 & 177 & 328 \\
\tableline
\end{tabular}					 
%\end{flushleft}
\label{t-compos}
\end{table}

%\clearpage

%\clearpage

\begin{figure}
%\centering
%\includegraphics[width=8cm]{f1.eps}
%%\epsscale{1.0}
\plotone{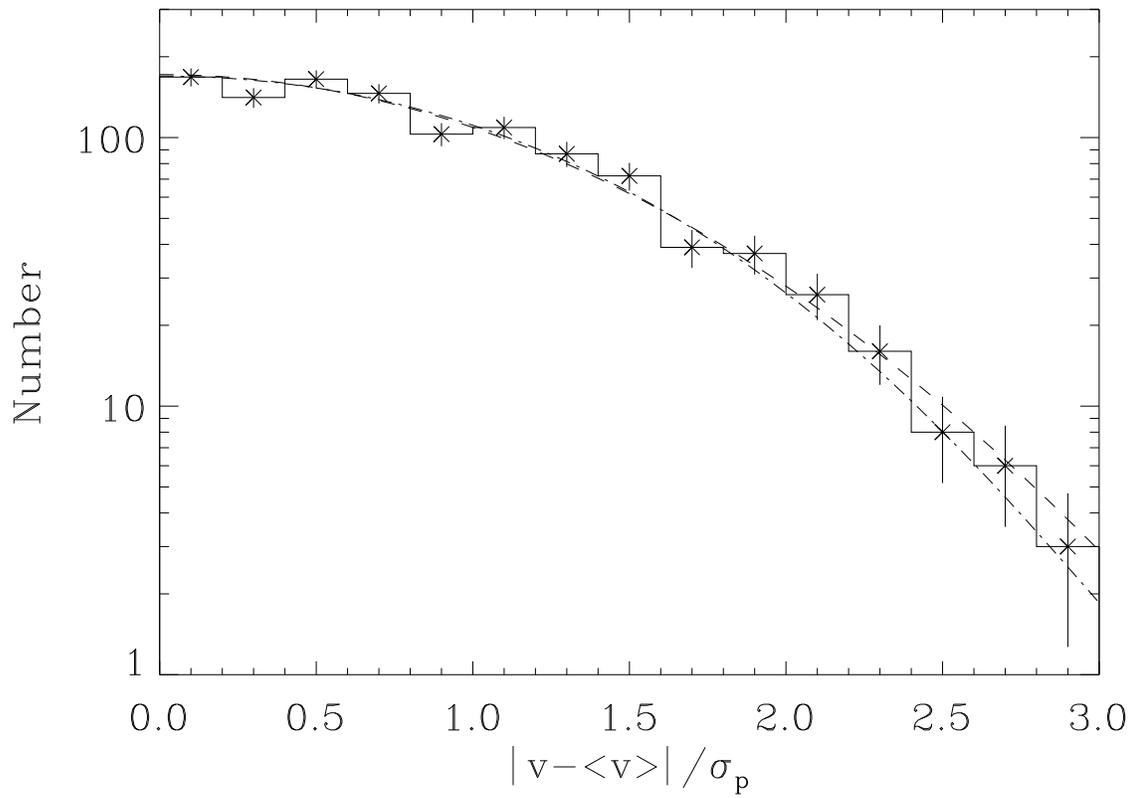}
\figcaption{The velocity distribution of the Early-type class, with the
best-fitting Gaussian (dashed-line), and the best-fitting Gauss-Hermite 
polynomial (dash-dotted line) superposed.
\label{f-gh}}
\end{figure}

%\clearpage

The analysis of the mass profile cannot be based on the sample of
galaxies within substructure because the members of substructure are
orbiting together, so that their kinematics does not only probe the
large-scale properties of the cluster potential. For the galaxies
outside substructure, it was found that there are 4 galaxy classes
that must be distinguished because they have different phase-space
distributions. These 4 classes are: (i) the brightest ellipticals
(with $M_R \leq -22+5 \log h$), which we refer to as '\ebr' in the
following, (ii) the other ellipticals together with the S0 galaxies
(referred to as 'Early-type' galaxies, in the following), (iii) the
early spirals (Sa--Sb), which we refer to as '\spe' in the following,
and (iv) the late spirals and irregulars (Sbc--Ir) together with the
ELG (except those with early morphology), globally referred to as
'\spl' in what follows.

In Table~\ref{t-compos} we show the number of cluster members outside
substructure in the 59 clusters, in each of the 4 galaxy classes
defined above. As explained in Appendix~\ref{a-prof}, in building the
number density profiles we need to correct for sampling
incompleteness. In order to keep the correction factor small, galaxies
located in poorly-sampled regions are not used in the present
analysis, and are not counted in Table~\ref{t-compos}.

\section{The Early-type galaxies as tracers of the mass profile}
\label{s-massearly}

Since we want to derive the mass profile of the ensemble cluster by
application of the Jeans equation of stellar dynamics, we must assess
the suitability of each of the 4 galaxy classes to serve as tracers of
the potential. To begin with, suitable tracers should be in
equilibrium with the cluster potential. However, application of the
Jeans equation also requires that the orbital characteristics are
known or, in other words: that we have information on the
(an-)isotropy of the velocity distribution.

For that reason, and in view of the results in papers III and VI, it
is very unlikely that the \spl~ can be used, since the analyses in
these papers suggest that they may be on first-approach orbits towards
the central regions. This could still mean that they are in
equilibrium with the potential, but as we have no {\em a priori}
knowledge of their orbits, except that they are unlikely to be
isotropic, these galaxies do not qualify as good tracers.

%\clearpage

\begin{figure}
%\centering
%\includegraphics[width=8cm]{f2.eps}
\epsscale{0.70}
\plotone{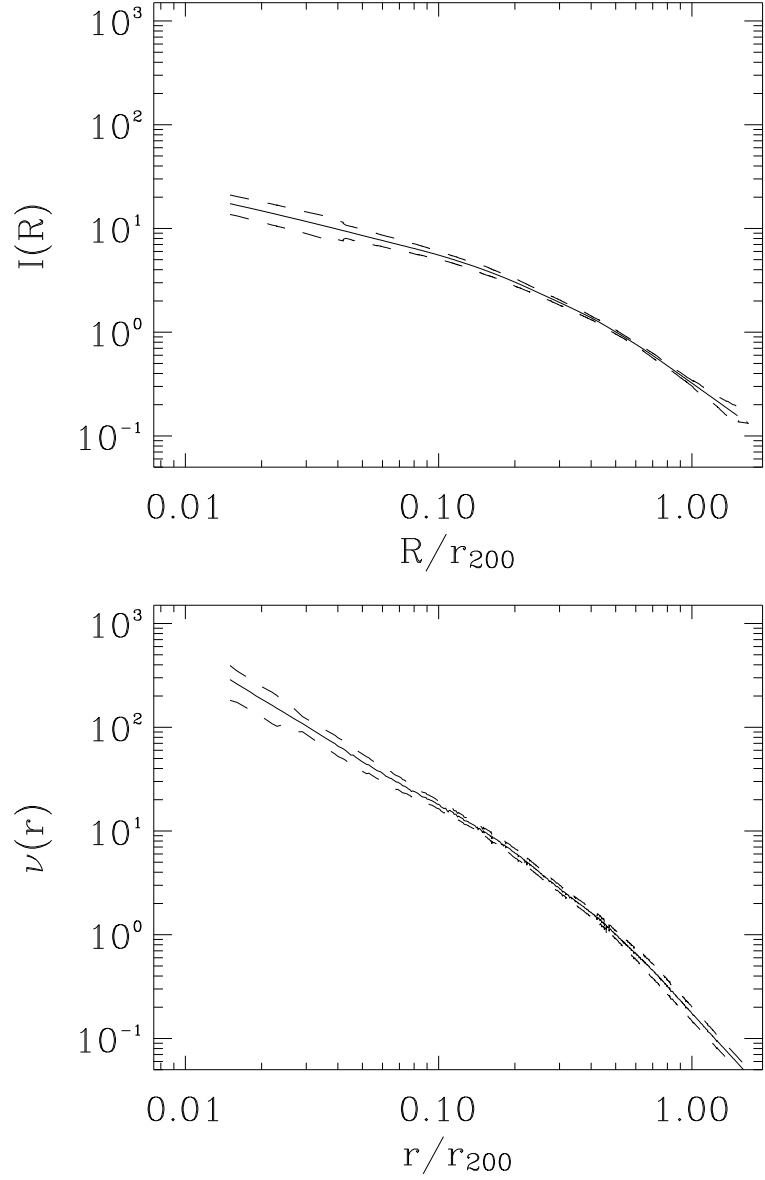}
\figcaption{Top: The best LOWESS estimate (solid line) of $I(R)$ of the 
Early-type galaxies, within the 1-$\sigma$ confidence interval
determined from bootstrap resamplings (dashed lines). Bottom: The best
estimate (solid line) of $\nu(r)$, extrapolated to $r=6.67 \,
r_{200}$, within the 1-$\sigma$ confidence interval determined from
bootstrap resamplings (dashed lines).
\label{f-dprof}}
\end{figure}

\begin{figure}
%\centering
%includegraphics[width=8cm]{f3.eps}
%%\epsscale{0.70}
\plotone{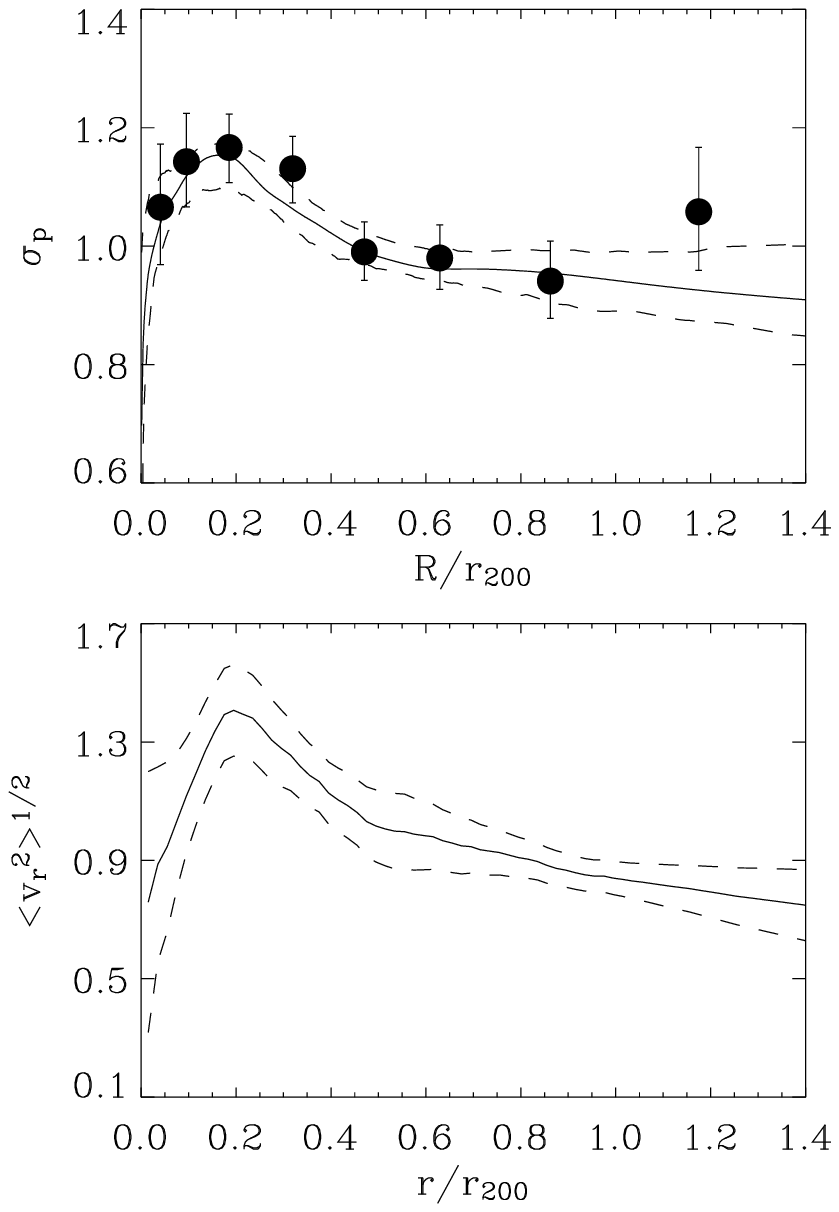}
\figcaption{Top: The best LOWESS estimate (heavy line) of $\sigma_p(R)$
of the Early-type galaxies, with 68\% confidence levels (dashed
lines), together with binned estimates. Bottom: The best estimate
(heavy line) of ${\rm{<} v_r^2 \rm{>}}(r)$, with 68\% confidence
levels. \label{f-vdp}}
\end{figure}

%\clearpage

The \ebr~ could, for all that we know, have an isotropic velocity
distribution. However, it is unlikely that, as a class, they satisfy
the basic assumption underlying the Jeans equation, namely that it
describes a collisionless particle fluid. The usual interpretation
that the brightest ellipticals have been directed to the central
regions through dynamical friction after which they have grown at the
expense of other galaxies might make them unfit for estimates of the
mass profile.

As to the choice between the two remaining classes, the relative
statistical weights clearly point to the use of the Early-type
galaxies as tracers of the mass profile. However, in addition to this
pragmatic argument there are other, more fundamental reasons for this
choice. The most important one is that the stellar population in most
ellipticals is quite old, and evidence is mounting that most of the
ellipticals formed before they entered the cluster. This should have
allowed them to settle in the potential and become good tracers of it.

As was shown in paper XI, the $(R,\rm{v})$-distribution of the S0
galaxies is very similar to that of the ellipticals. At first, this
may appear somewhat surprising. Although the stellar population of
many S0 galaxies is probably as old as that of the ellipticals, it is
generally believed that a significant fraction of S0's has formed
rather recently through the transformation of early spirals by
impulsive encounters. Apparently, the present rate of formation of
S0's is sufficiently low that the bulk of the transformation of early
spirals has taken place sufficiently long ago, so that the phase space
distribution of the S0's has relaxed to that of the older population
of E's.

Use of the Early-type galaxies requires that we know, or can make an
educated guess about, the anisotropy of their velocity distribution.
We will assume that their orbits are isotropic, and this assumption
can be checked to some extent because the distribution of relative
radial velocities depends on the anisotropy of the 3-D velocity
distribution, as was shown by Merritt (1987) and van der Marel et
al. (2000). By calculating the even-order Gauss-Hermite coefficients
for the velocity distribution of the CNOC1 survey, van der Marel et
al. (2000) showed that they could constrain the range of allowed
values of the velocity anisotropy, from a comparison with dynamical
models.

The velocity distribution of the Early-type galaxies in our ensemble
cluster is shown in Fig.~\ref{f-gh}, with the best-fitting Gaussian
and the best-fitting Gauss-Hermite polynomial superposed. It appears
that the deviations from a Gaussian are very small. This is confirmed
by the values of $h_4$ and $h_6$, which are $-0.016$ and
$0.005$. Although we refrain from constructing plausible distribution
functions, the projected number density -- and therefore also the 3-D
number density -- is sufficiently close to that used by van der Marel
et al. (2000) that we can conclude from their Fig.~8 that our
assumption $\beta(r) = 1 - {\rm{<}v_t^2 \rm{>}}(r) / {\rm{<} v_r^2
\rm{>}}(r) \equiv 0$ for the E+S0 class is very plausible (as a matter
of fact we conclude that $-0.6 \la \beta \la 0.1$, or, equivalently,
$0.8 \la \sqrt{\rm{<} v_r^2 \rm{>}/\rm{<} v_t^2 \rm{>}} \la 1.05$).

\section{Direct solution of the Jeans equation}
\label{s-direct}

Application of the Jeans equation for the determination of the cluster
mass profile requires two observables: the projected number-density
profile $I(R)$, and the velocity-dispersion profile $\sigma_p(R)$ of
the Early-type galaxies. Here we summarize the steps involved in the
determination of these profiles and their inversion to the 3-D
space. Details can be found in Appendix~\ref{a-prof}.

For the determination of both number-density and velocity-dispersion
profiles we use the LOWESS technique (Gebhardt et al. 1994). The
number-density profile, $I(R)$, is corrected for sampling
incompleteness. The confidence levels (c.l. hereafter) on $I(R)$ are
measured via a bootstrap procedure. The projected number-density
profile $I(R)$ is shown in the upper panel Fig.~\ref{f-dprof}. The
projected number density $I(R)$ must be deprojected to yield the 3-D
number density $\nu(r)$. This deprojection is obtained via a
straightforward integration of the Abel integral, which involves no
assumptions, apart from the behaviour of $I(R)$ towards large
radii. We checked that the deprojected profile is essentially
independent from how we extrapolate $I(R)$ to large radii (see
Appendix~\ref{a-prof}). We show the deprojected profile $\nu(r)$ in
the botttom panel of Fig.~\ref{f-dprof}.

In the upper panel of Fig.~\ref{f-vdp} we show the projected
velocity-dispersion profile, determined with the LOWESS technique,
together with a binned version, where the value of the velocity
dispersion in each radial bin is computed using the robust bi-weight
estimator (see Beers, Flynn, \& Gebhardt 1990).

With the estimates of $I(R)$, $\nu(r)$ and $\sigma_p(R)$ we now
proceed to obtain the cluster mass profile $M(<R)$.  The procedure is
well-known (see, e.g., BT). We make the usual assumptions that the
cluster is in steady-state, and that net rotation is negligible. With
these assumptions, and in the case of spherically symmetry (see also
Appendix~\ref{a-assum}), the procedure is as follows. First, one must
deproject $\sigma_p(R)$ to obtain ${\rm{<} v_r^2 {>}}(r)$. In other
words, one should solve the integral relation:
\begin{equation}
I(R) \sigma_p^2(R) = 2 \int_{R}^{\infty} \left( 1 - \beta(r)\frac{R^2}{r^2}
\right) \frac{r \nu(r) {\rm{<} v_r^2 {>}}(r) \, dr}{\sqrt{r^2-R^2}},
\label{e-abel2d}
\end{equation}
where, as before:
\begin{equation}
\beta(r) \equiv 1 - \frac{{\rm{<} v_t^2 \rm{>}}(r)}{{\rm{<} v_r^2 \rm{>}}(r)},
\label{e-beta}
\end{equation}
and ${\rm{<} v_r^2 \rm{>}}(r)$, ${\rm{<} v_t^2 \rm{>}}(r)$ are the mean
squared components of the radial and tangential velocity. For the special
case $\beta(r) \equiv 0$ the solution is:
\begin{equation}
{\rm{<} v_r^2 \rm{>}}(r) = - \frac{1}{\pi \nu(r)} \int_{r}^{\infty}
\frac{d [I(R) \times \sigma_p^2(R)]}{dR} 
\frac{dR}{\sqrt{R^2-r^2}}.
\label{e-abel3d}
\end{equation}
The ${\rm{<} v_r^2 \rm{>}}(r)$ profile and c.l. resulting from $I(R)$,
$\sigma_p(R)$, and the c.l. on $\sigma_p(R)$ (see
Appendix~\ref{a-prof} for details) are shown in the lower panel of
Fig.~\ref{f-vdp}.

The mass profile then follows directly from the Jeans equation (see,
e.g., BT):
\begin{equation}
M(<r) = - \frac{r {\rm{<} v_r^2 \rm{>}}}{G} \left( \frac{d
\ln \nu}{d \ln r} + \frac{ d \ln {\rm{<} v_r^2 \rm{>}}}{d \ln r} \right),
\label{e-jeans}
\end{equation}
where $\beta$ was assumed to be 0 (see \S~\ref{s-massearly}).

\begin{figure}
%\centering
%\includegraphics[width=9cm]{f4.eps}
%%\epsscale{0.55} 
\plotone{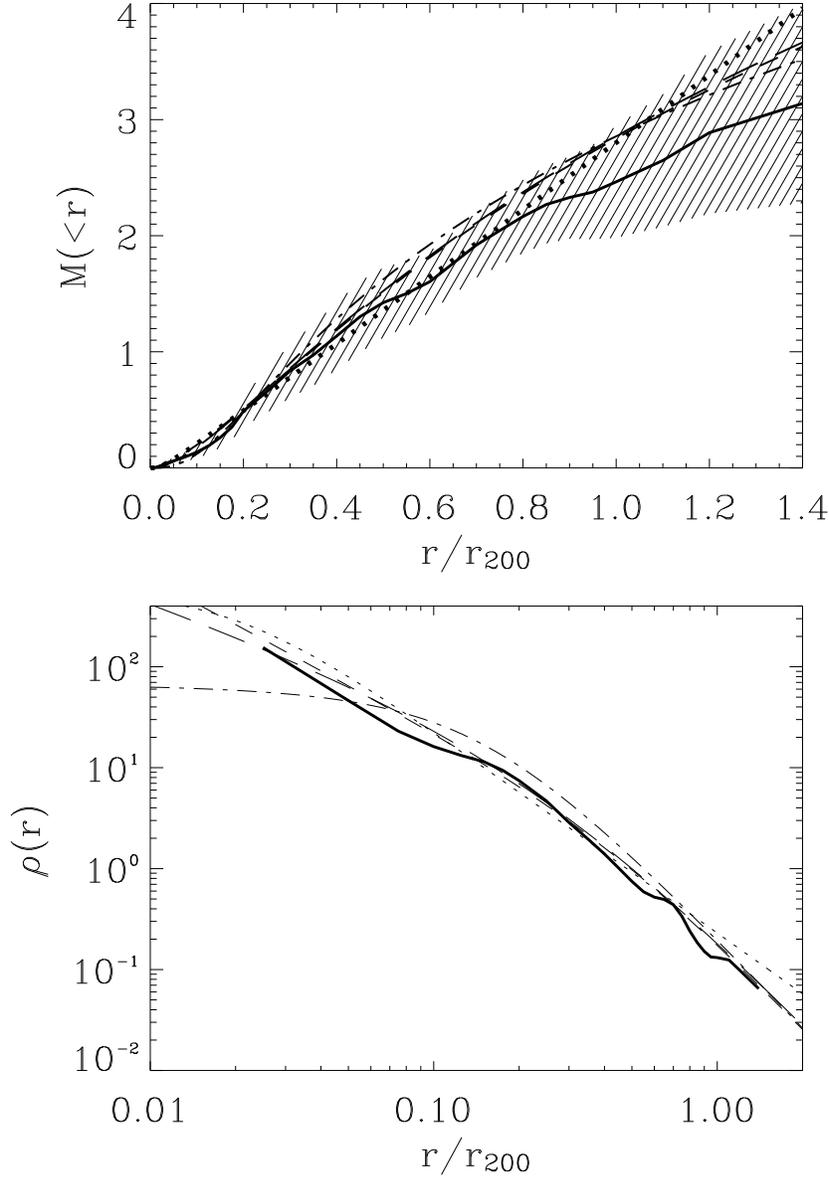} \figcaption{Top: The mass
profile $M(<~r)$ calculated from the Jeans equation, using the
Early-type galaxies as tracers of the potential (heavy solid
line). Isotropic orbits were assumed, and the approximate 68\%
confidence region is indicated by the shading. The mass scale is in
arbitrary units. Also shown are the best-fit mass models of the NFW
(long dashes), M99 (short dashes), SIS (dotted line) and Burkert
(dash-dotted line) types. Note that the best-fit mass models were not
derived from fits to $M(<r)$, but from a comparison of the observed
and predicted velocity-dispersion profiles (see \S~\ref{s-model}).
Bottom: the mass density profile $\rho(r)$ derived by differentiating
our observed $M(< r)$ (heavy solid line) compared with the 4 models
(same coding as above).\label{f-massprof}}
\end{figure}

The result is shown in the upper panel of Fig.~\ref{f-massprof}.
Since in our ensemble cluster galaxy velocities are normalised by
their cluster global velocity dispersion, $\sigma_p$, and galaxy
clustercentric distances are scaled by their cluster virial radius,
$r_{200}$, the mass is also expressed in normalised
units\footnote{Since the median velocity dispersion of our cluster
sample is 699 km~s$^{-1}$, and the median value of $r_{200}$ is 1.2 $
h^{-1} Mpc$ one mass unit corresponds to $\approx 1.4 \times 10^{14}
h^{-1} M_{\odot}$}. Also shown are the mass profiles of four popular
models: viz. those of NFW, M99, the softened isothermal sphere (SIS,
hereafter; see, e.g., Geller et al. 1999), and that of Burkert
(1995). They are all one-parameter models in which the linear scale
(in units of $r_{200}$) is variable. The analytic expressions are:
\begin{equation}
\rho_{NFW}(r) = \frac{\rho_0}{(r/r_s) (1+r/r_s)^2}
\end{equation}
\begin{equation}
\rho_{M99}(r) = \frac{\rho_0}{(r/r_M)^{1.5} [1+(r/r_M)^{1.5}]}
\end{equation}
\begin{equation}
\rho_{SIS}(r) = \frac{\rho_0}{1+(r/r_c)^2}
\end{equation}
\begin{equation}
\rho_{Burkert}(r) = \frac{\rho_0}{{(1+r/r_0)}{[1+(r/r_0)^2]}}
\end{equation}

Our mass profile $M(< r)$ and the four model profiles cannot be
compared directly, because the c.l. of our mass profile are only
approximate (as discussed in Appendix~\ref{a-prof}). For a proper
statistical comparison we need to project our mass profile in the
space of observables, and this is described in the next section. Yet,
for the purpose of illustration, we show in the lower panel of
Fig.~\ref{f-massprof} the mass density profile as derived by
differentiation of the observed mass profile $M(< r)$, together with
the four model density profiles. In both panels of
Fig.~\ref{f-massprof} we have used the best-fit values
$r_s/r_{200}=0.25$, $r_M/r_{200}=0.5$, $r_c/r_{200}=0.02$, and
$r_0/r_{200}=0.15$ (see below).

As a test of our assumption about the unsuitability of the late-type
galaxies as tracers of the potential (see \S~\ref{s-massearly}), we
have also applied the procedure, described here for the Early-type
galaxies, to the early and late spirals (again, assuming isotropy).
The \spe\ mass profile is not very different from that in
Fig.~\ref{f-massprof}, but the \spl\ mass profile is considerably
steeper, which supports our assumption. In a forthcoming paper
(Biviano \& Katgert, 2004), we discuss in detail the dynamical
equilibrium (including the orbital anisotropies) of the \ebr, \spe\
and \spl\ classes, in the potential derived here from the data of the
Early-type galaxies.

\section{Comparison with model mass-profiles}
\label{s-model}

In order to compare our observed mass profile with models from
the literature, we work in the domain of observables, viz. in
$\sigma_p(R)$. This is possible because one can solve for $\rm{<}
v_r^2 \rm{>} (r)$ using the observed $\nu(r)$, an assumed mass profile
$M(<r)$, and an assumed $\beta(r)$, as follows:
\begin{equation}
\nu(r) {\rm{<} v_r^2 \rm{>}}(r) = - G \int_{r}^{\infty} \nu(\xi)
\frac{M(<\xi)}{\xi^2} \, \exp \left[ 2 \int_{r}^{\xi} \frac{\beta 
dx}{x}\right] d \xi
\label{e-vdm}
\end{equation}
This solution, given by van der Marel (1994), is a special case of a
more general solution of the Jeans equation which was developed by
Bacon, Simien, \& Monnet (1983) for building dynamical models of elliptical
galaxies.  For $\beta=0$ the above equation reduces to:
\begin{equation} 
\nu(r) {\rm{<} v_r^2 \rm{>}}(r) = -G \int_{r}^{\infty}
\nu(x) M(<x) x^{-2} dx
\label{e-vdm0}
\end{equation}

Given a model mass profile, $M(<r)$, and an observed number-density
profile, $\nu(r)$, it is thus possible to compute model projected
velocity-dispersion profiles, through eq.~\ref{e-vdm0} and the usual
Abel relation.

%\clearpage
\begin{figure}
%\centering
%\includegraphics[width=9cm]{f5.eps}
\epsscale{1.0} 
\plotone{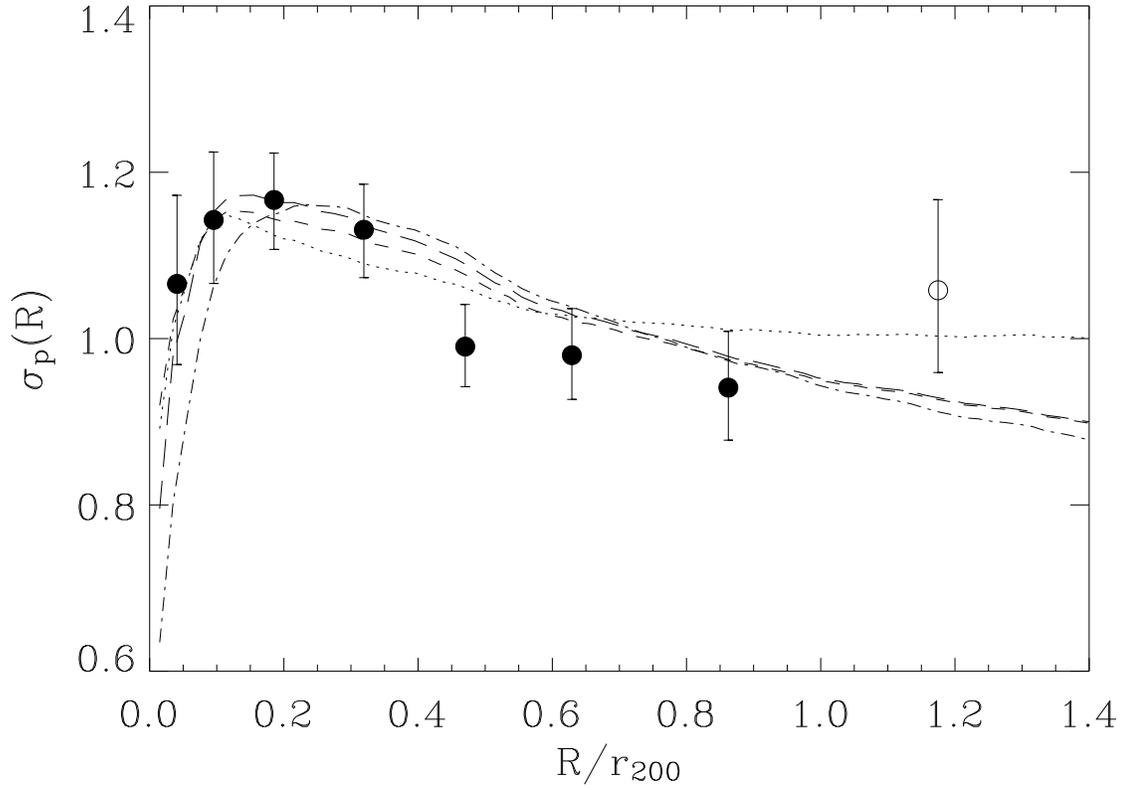} 
\figcaption{The observed
velocity-dispersion profile of the Early-type galaxies (filled circles
with error bars) and the four velocity-dispersion profiles predicted
for the 4 mass models: NFW (long dashes), M99 (short dashes), SIS
(dotted line) and Burkert (dash-dotted line). Note that the fits do
not include the outermost point (open circles).\label{f-vdpmodel}}
\end{figure}

The comparison between the model and the observed velocity-dispersion
profile yields a value for $\chi^2$ by which we measure the
acceptability of the assumed mass-profile model. A straightforward
determination of the $\chi^2$ value requires the observed profile to
have independent data-points. In this comparison we therefore use a
binned $\sigma_p(R)$-profile, rather than the LOWESS profile. Only the
data points within the virial radius $r_{200}$ are considered for the
fit, since galaxies at larger radii might not yet have relaxed to
dynamical equilibrium. The uncertainty in the observed $\nu(R)$,
determined from the bootstrap resamplings of $I(R)$ (see
Appendix~\ref{a-dprof}) is taken into account in the
$\chi^2$-analysis. In practice, for each bootstrap resampling of
$I(R)$ we have a corresponding $\nu(r)$, and hence, a different model
velocity dispersion profile, obtained via eq.~\ref{e-vdm0}. This
allows us to compute an r.m.s. for each point of the model velocity
dispersion profile. The model r.m.s. is then added in quadrature to
the uncertainty of the observed $\sigma_p(R)$ to give the full
uncertainty to be used in the $\chi^2$-analysis.

We checked that our results are robust w.r.t. different choices of
both the binning radii for $\sigma_p(R)$, and the way in which $I(R)$
is extrapolated to large radii to yield $\nu(r)$ through the Abel
inversion (see Appendix~\ref{a-prof}). Different choices of the
$\sigma_p(R)$ binning and of the $I(R)$ extrapolation affect the
best-fit values of the mass model parameters by $\la 15$\% and $\la
10$\%, respectively. 

We consider two 'cuspy' model mass profiles (NFW and M99), and two
profiles with a `core' (SIS and that of Burkert 1995) -- see
\S~\ref{s-massearly}. All four model mass profiles provide acceptable
fits to the observed $\sigma_p(R)$, although the models of NFW and
Burkert provide marginally better fits than the M99 and, in
particular, the SIS models. More specifically, using the $\chi^2$
statistics, we estimate the rejection probabilities of the best-fit
NFW, M99, SIS and Burkert models to be 19\%, 28\%, 72\% and 11\%,
respectively. Note that the $\chi^2$ values were calculated for the
points with $R/r_{200} < 1.0$, because inclusion of data beyond the
virial radius does not seem very sensible. As a mattter of fact, if
the outmost point is included all 4 models provide fits of similar
quality.

The 68\% c.l. range for the NFW-profile scale parameter is $0.15 \leq
r_s/r_{200} \leq 0.40$, with a best-fit value of $r_s=0.25 \,
r_{200}$. In terms of the concentration parameter, the best-fit value
is $c \equiv r_{200}/r_s =4_{-1.5}^{+2.7}$.  The 68\% c.l.  range for
the M99-profile scale parameter is $0.30 \leq r_M/r_{200} \leq 0.75$
with a best-fit value of $r_M=0.45 \, r_{200}$. On the other hand, for
no value of the scale parameter $r_c$ is the SIS model acceptable at
the 68\% c.l. or better. The best-fit is obtained for $r_c=0.02 \,
r_{200}$ and values $r_c > 0.075 \, r_{200}$ are rejected at $>99$\%
c.l. The 68\% c.l. range for the Burkert scale parameter $r_0 $ is
$0.10 \leq r_0/r_{200} \leq 0.18$ with a best-fit value of $r_0=0.15
\, r_{200}$. Values $r_0 > 0.25 \, r_{200}$ are rejected at $>99$\%
c.l.  The allowed range for the Burkert scale parameter seems larger
than for the SIS scale parameter. Note, however, that the two scale
parameters do not have the same meaning.  If we define the core radius
as the clustercentric distance where the density falls below half its
central value, $r_{\rho_0/2}$, this corresponds to $r_c$ in the SIS
model, and to $\sim 0.5 \, r_0$ in the Burkert model.  We conclude
that the core models (and in particular the SIS-model) that fit our
data have such small core-radii that they very much resemble the two
core-less models.

In Fig.~\ref{f-vdpmodel} we show the observed (binned) $\sigma_p(R)$
of the Early-type class, together with the velocity dispersion
profiles predicted from the best-fit NFW, M99, SIS and Burkert $M(<r)$
models.  For the sake of clarity, we do not show the uncertainties in
the model velocity dispersion profiles related to the uncertainties in
$\nu(r)$. These are however much smaller than the uncertainties of the
observed velocity dispersion profile. From Fig.~\ref{f-vdpmodel} it
can be seen that the best-fit NFW and M99 models predict almost
indistinguishable velocity dispersion profiles. On the other hand,
fitting the inner part of $\sigma_p(R)$ requires such a small core
radius for the SIS model, that the model velocity dispersion profile
flattens already at $r \approx 0.4 \, r_{200}$, while the observed
profile continues to drop. Formally, the best-fit Burkert model also
provides the best fit to the observations, as it reproduces the broad
maximum in $\sigma_p(R)$ better than do the other three models.
However, we are somewhat suspicious of the very strong decrease of
$\sigma_p(R)$ below $r \approx 0.1 \, r_{200}$, although it is only
$2.5 \, \sigma$ lower than the LOWESS estimate of $\sigma_p(R)$.

\section{The radial variation of the mass-to-light ratio 
         ${\cal M}/{\cal L}$}
\label{s-mlprof}

As mentioned in the Introduction, there is not yet a clear,
unambiguous result for the dependence of ${\cal M}/{\cal L}$ on radius
in the literature, and therefore we have also used our data to derive
${\cal M}/{\cal L}(r)$. The luminosity-density profile ${\cal
L}_{all}(r)$ was determined from the projected luminosity density
profile of the galaxies with the same method used to deproject $I(R)$,
except that galaxies are now weighed by their luminosities. Note that
the luminosity density profile is constructed for all galaxies, both
in and outside substructures. As discussed in Appendix~\ref{a-prof},
${\cal L}_{all}(r)$ only accounts for the luminosities of the galaxies
in our sample, as we did not try to recover the luminosity of unseen
cluster galaxies (e.g., by fitting a luminosity function to the
observed luminosity distribution, and extrapolating beyond the
completeness limit). In fact, we are not interested in estimating the
absolute values of ${\cal M}/{\cal L}$, but only in its radial
behaviour, and there is no way we could estimate ${\cal L}(r)$ for the
total galaxy population, without actually measuring positions and
luminosities of all members. The full magnitude range of the galaxies
on which ${\cal L}_{all}(r)$ is based is $-23.3 \leq M_R-5 \log h \leq
-17.4$, with an average $<M_R>=-20.25 + 5 \log h$, and an r.m.s. of
$0.94$ mag. The average absolute magnitude is about one magnitude
fainter than the knee of the cluster galaxy luminosity function, $M^*$
(e.g. Yagi et al. 2002). Note that our magnitudes are K-corrected, but
that we did not apply any correction for evolutionary effects. For
redshifts $\la 0.1$, those are very small in $R$-band, and the {\em
differential} evolutionary correction for early- and late-type
galaxies is estimated to be less than 5\% in luminosity.

\begin{figure}
%\centering
%\includegraphics[width=9cm]{f6.eps}
\epsscale{0.7}
\plotone{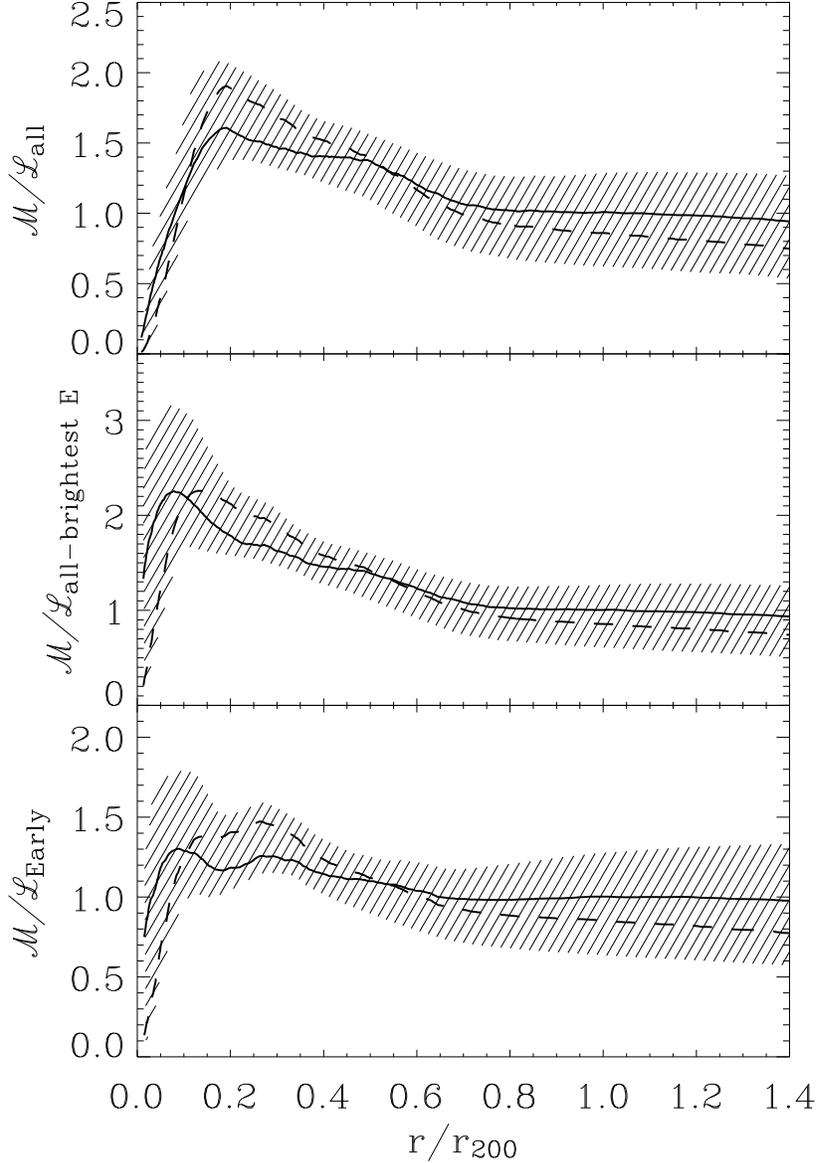}
\figcaption{Top: ${\cal M}/{\cal L}(r)$-profile, normalized at $r_{200}$, 
and calculated as $\rho(r)/{\cal L}_{all}(r)$ for all galaxies in our
sample. ${\cal L}_{all}(r)$ is the deprojected 3-D luminosity profile
and $\rho(r)$ corresponds to the best-fitting NFW (solid line) and
Burkert (dashed line) models. The shaded region denotes the 68\%
uncertainties in the profile ratio, as calculated from the union of
the 68\% c.l. ranges of those models. Center: ${\cal M}/{\cal
L}(r)$-profile for all galaxies except the brightest ellipticals,
expressed in normalized units and calculated as $\rho(r)/{\cal L}_{all
- brightest \, E}(r)$. The 68\% uncertainties in the profile ratio are
calculated as described above. Bottom: ${\cal M}/{\cal L}(r)$-profile,
expressed in normalized units and calculated as $\rho(r)/{\cal
L}_{Early}(r)$ for the Early-type galaxies only. The 68\%
uncertainties in the profile ratio are calculated as described above.
\label{f-lnfw}}
\end{figure}

In the upper panel of Fig.~\ref{f-lnfw} we show the ratio of the total
mass-density profile $\rho(r)$ of the best fitting NFW and Burkert
models, and ${\cal L}_{all}(r)$ calculated for all galaxies, i.e.  the
differential mass-to-light ratio as a function of $r$. The shaded
region shows the 68\% uncertainties in the profile ratio, as
calculated from the union of the 68\% c.l. ranges of the best-fitting
NFW and Burkert models. Note that the relative uncertainties in ${\cal
L}_{all}(r)$ are much smaller than the allowed ranges of $\rho(r)$.
The ${\cal M}/{\cal L}_{all}(r)$-ratio is consistent with a constant
value in the radial range $0.2 \leq r/r_{200} \leq 1.4$, even though
there is an apparent decrease from $0.2 \, r_{200}$ to $0.7 \,
r_{200}$.

However, within $0.1 \, r_{200}$ the ${\cal M}/{\cal L}_{all}$-ratio
is significantly lower than at larger radii. The reason for that could
obviously be that the \ebr \, which are much more centrally
concentrated than the other galaxy classes, contribute relatively
little mass for their luminosity. To check this explanation, we also
determined the ${\cal M}/{\cal L}$-ratio for all galaxies without the
\ebr \, (i.e.  ${\cal M}/{\cal L}_{all - brightest \, E}(r))$. The
result is shown in the middle panel of Fig.~\ref{f-lnfw}.  As
expected, the strong decrease of the ${\cal M}/{\cal L}$-ratio within
$0.2 \, r_{200}$ has largely disappeared. However, the apparent
decrease of ${\cal M}/{\cal L}$ from $0.2 \, r_{200}$ to $0.7 \,
r_{200}$ is now stronger.

To enable comparison with data on high-redshift clusters, where
selection of cluster members is often done using the red-galaxy
sequence in the colour-magnitude diagram, we also determined the
${\cal M}/{\cal L}$-ratio for the Early-type galaxies only (again
excluding the \ebr); the result is shown in the lower panel of
Fig.~\ref{f-lnfw}. The apparent decrease of ${\cal M}/{\cal L}$ from
$0.2 \, r_{200}$ to $0.7 \, r_{200}$ has disappeared, and the ${\cal
M}/{\cal L}$-ratio for the Early-type galaxies only, is consistent
with a constant value. In other words: the luminosity of the
Early-type galaxies traces the total mass very well.  Therefore, we
conclude that to first order, the ${\cal M}/{\cal L}(r)$-profile
can be understood as a flat profile due to the Early-type galaxies,
which is modified by the 'extra' luminosity of the \ebr~ within $0.2
\, r_{200}$, and by the luminosity of the spirals beyond $0.2 \,
r_{200}$.

\section{Discussion}
\label{s-disc}

\subsection{The Mass Profile}

Using the ENACS data-set complemented with recently obtained
morphological data, we determined the mass and mass-to-light profile
for an 'ensemble' cluster built from 59 nearby clusters. Following the
results of the analysis of paper XI, the sample of
cluster galaxies was divided into four classes, viz. the brightest
ellipticals (\ebr), the Early-type galaxies (the other ellipticals and
the S0 galaxies), and the early and late spirals. We used the
Early-type class to obtain a non-parametric estimate of the mass
profile through the application of the Jeans equation (see
\S~\ref{s-direct}), and we fitted mass-profile models (see
\S~\ref{s-model}). In these analyses, we assumed isotropic orbits for
the Early-type galaxies, which is supported by a Gauss-Hermite
polinomial decomposition of their velocity distribution.

Both `cuspy' models, like the NFW and M99 mass models, and models with
a core, like the SIS and that proposed by Burkert (1995), provide
acceptable fits to the data. This follows from a comparison of the
observed and predicted velocity-dispersion profile of the Early-type
galaxies. However, the best-fit core models have such small core-radii
that they are not too different from the cuspy models. In other words,
we did not find compelling evidence for the existence of a 'core' in
the matter distribution in clusters.

This result, which applies to our 'ensemble' cluster is consistent
with several previous results based on smaller cluster samples, and
obtained by various methods. Among the latter are those based on an
analysis of the distribution and temperature of the X-ray emitting gas
(e.g.  Markevitch et al. 1999; Allen et al. 2002; Arabadjis, Bautz, \&
Garmire 2002; Pratt \& Arnaud 2002), those based on an analysis of the
weak-lensing distortion of background galaxies (e.g. Lombardi et
al. 2000; Clowe et al. 2000; Clowe \& Schneider 2001; Athreya et
al. 2002; Hoekstra et al.  2002; King et al. 2002; Dahle, Hannestad,
\& Sommer-Larsen 2003), and those based on an analysis of the
projected phase-space distribution of the galaxies (e.g. Geller et
al. 1999; van der Marel et al. 2000; Rines et al. 2002; Biviano \&
Girardi 2003). In some of these analyses, not only was consistency
with an NFW-type mass profile established, but an upper limit to the
size of a core could also be given (e.g., Arieli \& Rephaeli 2003;
Dahle et al. 2003). Yet, some clusters also show a sizeable core
(Shapiro \& Iliev 2000; Ettori et al. 2002; Sand et al. 2002).

Since most of these studies are based on small numbers of clusters, it
is possible (if not likely) that individual clusters have different
types of mass profile. If so, the result of our analysis is that {\em
on average} rich clusters are characterized by cuspy profiles, and
that, if clusters with cores do exist, they must be a minority. It is
noteworthy that the two previous analyses involving 'ensemble'
clusters, drawn from the CNOC and the 2dFGRS datasets respectively (by
van der Marel et al. 2000, and by Biviano \& Girardi 2003), give a
similar result. As a matter of fact, the best-fit value of the NFW
scale parameter obtained in these studies ($r_s/r_{200}=0.24$ and
$r_s/r_{200}=0.18$), is fully consistent with our value of
$r_s/r_{200}=0.25_{-0.10}^{+0.15}$.

Our result, that {\em on average} rich clusters have cuspy mass
profiles in the central region, is in agreement with predictions of
hierarchical Cold Dark Matter (CDM, hereafter) models. Our data do not
allow us to choose between different predictions, like the NFW and M99
models. Apart from the fact that the latter differ only slightly, the
differences are limited to the central region, where the (baryonic)
mass contribution of the cD is dominant (when present). Proper
comparison would require the analysis of the internal velocity
dispersion of the cD (Kelson et al. 2002). In any case, recent results
from numerical simulations suggest that the concept of a universal
inner slope of the mass profile might be inappropriate, as dark halo
mass profiles do not seem to converge to a single central power-law
shape (Power et al. 2003).

The allowed range $r_s/r_{200}=0.25_{-0.10}^{+0.15}$, that we obtain
for our ensemble cluster brackets the value predicted for a rich
cluster halo in a $\Lambda$CDM cosmology, $r_s/r_{200} \simeq 0.16$
(see Fig.6 in NFW). Note however, that we determine the profile of
{\em total} cluster mass, and not only of the {\em dark} mass. Our
result therefore includes the contribution of baryons, i.e. of
galaxies and the ICM. The latter can be a substantial fraction of the
total mass (e.g. Ettori, Tozzi, \& Rosati 2003). Since the ICM is less
centrally concentrated than the dark mass (see, e.g., Markevitch \&
Vikhlinin 1997; Pratt \& Arnaud 2002), one expects the total mass
distribution to be less concentrated than the dark mass. That might be
the reason why our best-fit NFW $r_s$ value is slightly larger than
the theoretical prediction (although not significantly so). On the
other hand, the baryonic mass contribution of the cD in the cluster
center has the opposite effect, namely of making the total mass
distribution {\em more} centrally concentrated than that of the dark
mass (Kelson et al. 2002; Dahle et al. 2003). However, separation of
the different components of the mass profile (although important) is
beyond the scope of this paper.

Cosmological simulations predict an increase of $r_s$ with the mass of
the system (NFW), and therefore a decrease of $r_s$ with redshift of
the system (Bullock et al. 2001). The predicted changes are rather
small; $r_s/r_{200}$ increases by only $\sim 25$\% over a decade in
cluster mass, while from $z=0$ to $z \approx 0.5$ it decreases by a
similar amount. Comparing our low-z, 68\% c.l. range of $r_s/r_{200}$
with the value obtained for the CNOC clusters studied by van der Marel
et al. (2000), of similar mass but at higher redshifts ($0.17 \leq z
\leq 0.55$), we find no redshift dependence of
$r_s/r_{200}$. Comparing our 68\% c.l. $r_s/r_{200}$-range with the
value of $\simeq 0.18$ for the poor clusters studied by Biviano \&
Girardi (2003), with masses 3 times smaller than ours, and with the
value of $\simeq 0.13$ for the groups studied by Mahdavi et
al. (1999), with masses 13 times smaller than ours, we find a trend
that is qualitatively in agreement with the theoretical prediction,
but which is not statistically significant.

The upper limit that we obtain for the size of a possible core
provides a constraint for the cross-section of Self-Interacting Dark
Matter (SIDM) models. SIDM models have become popular in recent years
as a means to explain the rotation curves of low surface brightness
galaxies (see, e.g., Firmani et al. 2000).  We translate the
constraints on the scale parameters of the SIS and Burkert models, by
using the results of the simulations of Meneghetti et al. (2001, see
their Table 1), and their definition of the core radius.  Our results
imply an upper limit to the scattering cross-section of dark matter
particles of $\sigma_{\star} < 1$ cm$^2$~g$^{-1}$ at $>99$\% c.l., in
good agreement with the conclusions of Meneghetti et al. (2001).

\subsection{The Mass-to-Luminosity Density Profile ${\cal M}/{\cal L}$(r)}

Using the magnitudes of the ENACS galaxies, together with the
best-fitting `cusp-' and `core-' mass models, we have studied the
radial variation of the ${\cal M}/{\cal L}$-ratio. When the luminosity
is based on all galaxies together, the ${\cal M}/{\cal L}$-ratio
increases strongly from $r = 0$ to $0.2 \, r_{200}$, decreases
somewhat from $0.2 \, r_{200}$ to $0.7 \, r_{200}$ and is constant
beyond $0.7 \, r_{200}$. The strong decrease to very low values of
${\cal M}/{\cal L}$ within $0.2 r_{200}$ is found to be largely due to
the \ebr, i.e.  the brightest ellipticals with $M_R \leq -22+5 \log
h$, because it disappears when the latter are excluded. Similarly, the
decrease of ${\cal M}/{\cal L}$ from $0.2 \, r_{200}$ to $0.7 \,
r_{200}$ largely disappears when the early and late spirals are
exluded. Thus, the total mass density is traced remarkably well by the
Early-type (elliptical and S0) galaxies. Although the Early-type
galaxies are the dominant population in present-day clusters, it does
not follow automatically that their luminosity should trace the total
mass so well.

As a matter of fact, the {\em virial} mass profile, estimated assuming
that the mass density is proportional to the number density of the
Early-type galaxies (see e.g. Girardi et al. 1998), generally
overestimates the mass, when compared with our solution in
Fig.~\ref{f-massprof} (by as much as a factor of 2, although it is
reduced, but does not disappear, when a correction is made for the
pressure term, see The \& White 1986). A similar virial-mass bias was
found by Carlberg et al. (1997b) for the CNOC clusters.

Several previous studies also yielded a flat or nearly-flat cluster
${\cal M}/{\cal L}$-profile when only early-type or red galaxies were
selected (e.g. Carlberg et al. 1997a; van der Marel et al. 2000; Rines
et al. 2001; Hoekstra et al. 2002; Biviano \& Girardi 2003; Ettori \&
Lombardi 2003). However, near the cluster center, where the inclusion
of the very bright galaxies in the sample lowers the cluster ${\cal
M}/{\cal L}$-value, the deviation from the flat profile was noted
before (see, e.g., Athreya et al. 2002). Outwardly decreasing ${\cal
M}/{\cal L}$-profiles are generally obtained if only late-type or blue
galaxies are considered (Carlberg et al. 1997b; Biviano \& Girardi
2003). However, using the red photometric band, Rines et al. (2000)
found a steeply decreasing ${\cal M}/{\cal L}$-profile in
A576. Currently, the ${\cal M}/{\cal L}$-profiles of groups is still
controversial, because it is difficult to derive a mass profile for
those systems (e.g. Mahdavi et al. 1999; Carlberg et al. 2001).

Two competing processes affect the relative distributions of mass and
luminosity in a cluster: viz. dynamical friction, and tidal stripping.
According to Mamon (2000) and Takahashi et al. (2002) the effect of
tidal stripping should dominate over the effect of dynamical friction,
so that massive galaxies sink toward the cluster center but loose
their halos due to tidal stripping. The result is an ${\cal M}/{\cal
L}$-value that increases towards the cluster center. This effect is
also observed in numerical simulations (Ghigna et al. 1998).
Luminosity segregation of the very massive galaxies can result at an
early epoch of the cluster formation (Governato et al. 2001). The fact
that we do not find significant segregation of the cluster ellipticals
and S0s with respect to the cluster mass (if we exclude the very
bright ellipticals) suggests that tidal stripping is able to remove
the dark haloes of these galaxies, but not to affect the galaxies
luminosities very much. If tidal stripping affected the luminosities,
of the Early-type galaxies, we should expect a decrease of their mean
luminosity towars the center, but our data do not show such a trend
(see paper XI). This is also consistent with the finding of Natarajan,
Kneib, \& Smail (2002b), based on a gravitational lensing analysis of
six galaxy clusters, that the typical halo truncation radius of
cluster galaxies is 5--10 times larger than the optical radius.

\section{Summary and conclusions}
\label{s-summ}

We have determined the mass profile of an 'ensemble' cluster built
from 59 nearby clusters from the ESO Nearby Abell Cluster Survey. In
this ensemble cluster, we distinguished four different classes of
galaxies, viz. the brightest ellipticals (\ebr, with $M_R \leq -22+5
\log h$), the Early-type galaxies, and early and late spirals, because
they have significantly different projected phase-space distributions.
Galaxies in substructures were excluded from the analysis.

Because the Early-type galaxies have a nearly isotropic velocity
distribution we derived a non-parametric estimate of the cluster mass
profile from their number-density and velocity-dispersion profiles,
through the Jeans equation of stellar dynamics. We compared our result
with models for the mass profile of dark matter halos, viz. the NFW,
M99, SIS and Burkert models. From a comparison of observed and
predicted velocity-dispersion profiles of the Early-type galaxies, it
appears that all 4 mass models provide acceptable fits to the
data. The best-fitt NFW model has a concentration parameter of
$c=4_{-1.5}^{+2.7}$ (68\% confidence level), in agreeement with
results from numerical simulations. The (SIS and Burkert) core-models
are only acceptable if their core-radius is sufficiently small 
(i.e. $r_{\rho_0/2} \la 0.13 \, r_{200}$ at the 99\% c.l.).

The ${\cal M}/{\cal L}$-ratio of the Early-type galaxies is remarkably
independent of radial distance. The ${\cal M}/{\cal L}$-ratio of all
galaxies together is not constant, and within $0.2 \, r_{200}$ it
shows a strong decrease towards the cluster center, which is due to
the \ebr. The spirals cause a mild decrease of the ${\cal M}/{\cal
L}$-ratio towards larger radii.

Our results support CDM models for the build-up of galaxy clusters,
and put an upper limit to the cross-section of any SIDM that could
dominate the cluster gravitational potential. We caution that our
results only apply to the total mass profile, since in our analysis we
cannot disentangle the baryonic and dark matter components.
Subtracting the ICM component would make the mass profile steeper, but
subtracting the contribution of the central cD (which is present in
many of the 59 clusters) would make it shallower.

\acknowledgments 
We thank Tom Thomas, Gary Mamon, and Srdjan Samurovi\'{c} for useful
discussions.  AB acknowledges the hospitality of the Leiden and
Marseille Observatories. PK acknowledges the hospitality of the
Trieste Observatory. This research was partially supported by the
Leids Kerkhoven-Bosscha Fonds, Leiden Observatory, and the Italian
Ministry of Education, University, and Research (through MIUR grant
COFIN2001028932 "Clusters and groups of galaxies, the interplay of
dark and baryonic matter").

\appendix

\section{The rejection of interlopers}
\label{a-interp}

%\clearpage

\begin{figure}
%\centering
%\includegraphics[width=9cm]{f7.eps}
\epsscale{1.0} 
\plotone{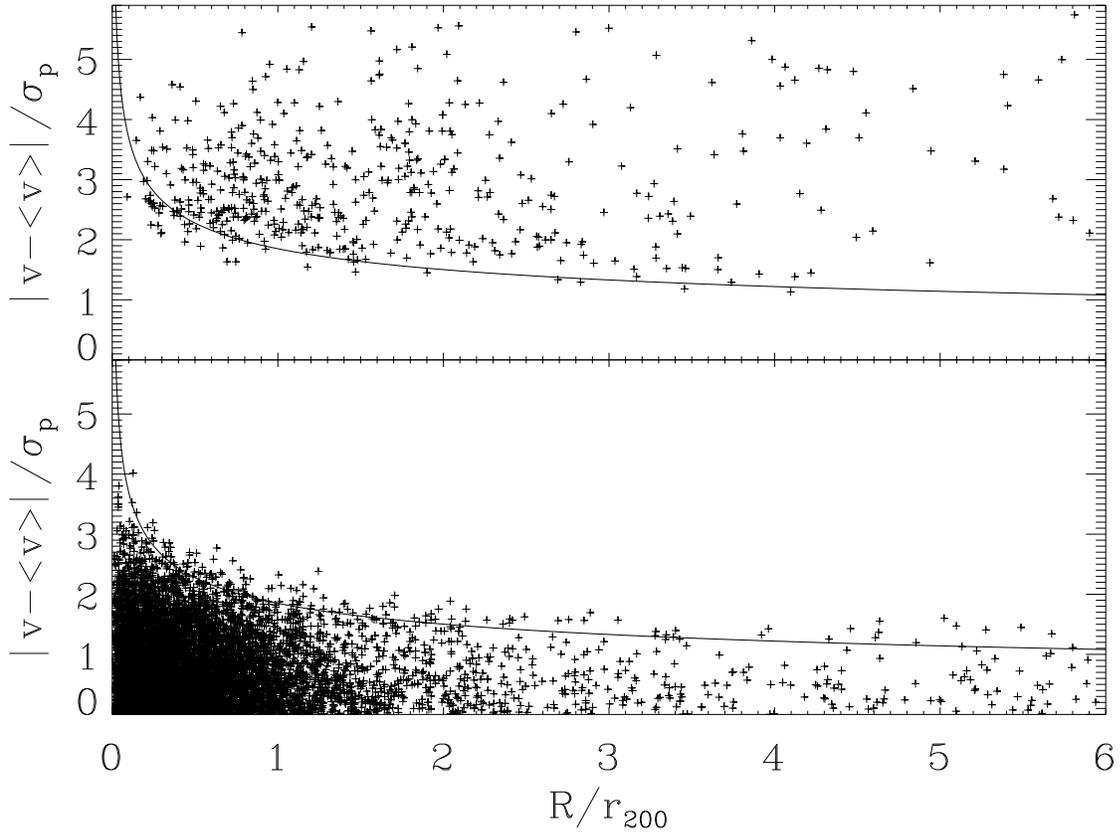} 
\figcaption{The distribution w.r.t. $R/r_{200}$ and $\mid
\rm{v}-{\rm{<} v \rm{>}} \mid/\sigma_p$ of the galaxies identified as
interlopers (top) and members (bottom) in 79 clusters with at least 45
galaxies with redshift.
\label{f-intlop}}
\end{figure}

%\clearpage

It is obviously very important that the analysis of the mass profile
is done on galaxy samples from which the non-members have been
removed. We indentify non-members (or interlopers) with the procedure
devised by den Hartog \& Katgert (1996). This procedure is based on an
iterative determination of an 'interim' mass profile, with which
galaxies that have relative line-of-sight velocities that are
inconsistent with their projected position and with the mass profile
can be found. The procedure is quite conservative in that it does not
eliminate cluster members, while galaxies outside the turn-around
radius can be recognized quite easily. The performance of the method
was checked on numerical models of galaxy clusters by van Kampen
(1994; see also van Kampen 1995) which describe a two-component
ensemble of galaxies and dark matter.

The procedure requires a minimum number of galaxies to start with,
because the interim mass profile needs to be sufficiently well defined
that the elimination of interlopers is meaningful. In practice, den
Hartog \& Katgert found that if the number of galaxies is less than
about 45, the iterative procedure quite often does not converge, while
for 45 galaxies or more the convergence generally is quite fast. We
applied the method to the clusters in our sample which have at least
45 galaxies with redshifts. For the clusters with less than 45
galaxies with redshifts, we used the interloper identification of the
'richer' clusters in a statistical sense.

Since the interloper rejection method is blind to galaxy morphologies,
the statistical definition of the separation between members and
interlopers does not require us to limit ourselves to clusters with
sufficient morphological data. Therefore we used 79 clusters with more
than 45 galaxies with redshifts, of which 32 are from ENACS and 47
from the literature. In the upper panel of Fig.~\ref{f-intlop} we show
the 671 interlopers identified with the method of den Hartog \&
Katgert (1996) in these 79 clusters, which contain a total of 8829
galaxies. At each projected radial distance $R$ there is a
well-defined lower limit to the velocity distribution of the
interlopers. In the lower panel of the same figure we show the 8158
galaxies that were accepted as member galaxies. In both panels, we
draw the same caustic that we visually defined as a good separation
between the two subsets.  Note that the inclusion of data from the
literature helps definining the caustic, as the ENACS data are mostly
restricted to the inner $\sim 1.5 \, r_{200}$.

The caustic defined in this way was used for the identification of the
interlopers in clusters with less than 45 galaxies. Because we
transfer the caustic in the \mbox{$(R/r_{200}, \mid \rm{v}-{\rm{<} v
\rm{>}} \mid/\sigma_p)$}-plane, the criterion for removing interlopers
is not systematically different for the two sets of clusters.

\section{The density and velocity dispersion profiles}
\label{a-prof}

\subsection{Number- and luminosity-density profiles}
\label{a-dprof}

We determine the projected number density profiles $I(R)$ for
the galaxy classes discussed in \S~\ref{s-data} as follows.
Because the Jeans equation requires the logarithmic derivative of the
deprojected number density profile, $\nu(r)$, a smooth representation
of $\nu(r)$ is needed. This is most easily obtained if the estimate of
$I(R)$ is also smooth, i.e. not binned. In order to obtain a smooth
estimate of $I(R)$ we have developed a variant of the LOWESS
technique, which was originally designed by Gebhardt et al. (1994) to
produce smooth velocity dispersion profiles.

The galaxies are ordered in increasing projected distance $R$. At the
position of each galaxy the density $I(R)$ is calculated as the
inverse of the distance between the n-th neighbours on either side.  A
density estimate typically involves about 10 galaxies. A smoothed
version of these densities is produced with the LOWESS routine,
through a weighted linear fit at each galaxy position, which uses
between 30 and 80\% of all densities, but with a weight that steeply
drops at large distances from the galaxy in question. We chose the
LOWESS smoothing factor guided by the binned versions of $I(R)$.

In the estimation of $I(R)$, each galaxy has a weight associated. This
is because the ENACS spectroscopy was done with circular plug-plates
used to position the fibers. Even if a cluster was observed with a
single plate, this was not necessarily centered on what, after the
spectroscopy, turned out to be the cluster center. The resulting
azimuthal incompleteness, which at large projected distances can
actually be quite severe (see e.g. Fig.~10 in paper I), depends only
on the positioning of the plate(s) with respect to the cluster center,
and can be easily calculated. Assuming azimuthal symmetry, which
should be a good approximation for the ensemble cluster, we derived
this so-called Optopus weight, which is the inverse of the fraction of
the azimuth covered in the parent cluster of the galaxy, at its
projected distance. Galaxies beyond the distance where the Optopus
weight gets larger than 3.33 are ignored in the analysis of the mass
profile.

The Optopus weight was multiplied by a second weight, which takes into
account the fact that in the ensemble cluster, not all clusters do
contribute at large projected distances $R/r_{200}$. If not properly
accounted for, this incompleteness effect would produce an artificial
steepening at larger radii. The correction for this effect assumes
that in the radial range where a cluster does not have data, its
contribution can be `invented' from the clusters that {\em do} have
data, as first described by Merrifield \& Kent (1989). In our
application of the method we invent data only from measured data, and
not from data that are themselves (partly) invented.

We check the consistency of the smoothed $I(R)$ with 64 bootstrap
resamplings of the data, because the latter give a good idea of the
range of $I(R)$'s implied by our data. The results are shown in
Fig.~\ref{f-dprof} (upper panel). We show the best LOWESS estimate
(heavy line), within the 68\% confidence interval determined from the
64 bootstrap resamplings (dashed lines).

We determine the deprojected profile , $\nu(r)$, for each of the 64
bootstrap resamplings of $I(R)$, through the standard Abel
deprojection equation (see, e.g., BT), viz.
\begin{equation}
\nu(r) = - \frac{1}{\pi} \int_{r}^{\infty} \frac{dI}{dR} 
\frac{dR}{\sqrt{R^2-r^2}} 
\label{e-abel1}
\end{equation}
via numerical integration. The deprojection requires knowledge of the
$I(R)$ to arbitrarily large $R$. In practice, we set the upper limit
of the integral equal to $R_t \equiv 6.67 \, r_{200}$ (it corresponds
to $\sim 10 h^{-1}$~Mpc for a massive cluster), since this is
sufficiently far from the last measured point. We use the following
smooth parametric function for the extrapolation of $I(R)$,
\begin{equation}
I(R \geq R_l)=C \frac{(R_t-R)^a}{R^b}.
\label{e-irext}
\end{equation}
where $R_l$ is the largest value of $R$ among the galaxies in the
sample. Continuity of $I(R)$ and its derivative at the last measured
point, $R_l$, fixes the values of the parameters $C$ and $b$, while
the shape of the extrapolation depends on the value of $a$. However,
it turns out that for $0.5 < a < 2.0$ the detailed form of the
extrapolation is not important, as different values of $a$ lead to
changes in $\nu(r)$ which, when projected, produce variations in
$I(R)$ which are much smaller than the intrinsic uncertainties (as
indicated by the bootstrap resamplings).

The deprojected number density profile so obtained is shown in
Fig.~\ref{f-dprof} (lower panel) for $a=1$. We show the best estimate
(heavy line) within the 1-$\sigma$ confidence interval determined from
the 64 bootstrap resamplings (dashed lines).

The same procedure is adopted for the determination of the
luminosity-density profiles, ${\cal L}(r)$, but with an additional
weight proportional to the luminosity of each galaxy, $10^{-0.4 M_R}$,
where $M_R$ is the absolute magnitude. In Paper V we have shown that
the ENACS galaxy samples form unbiased subsets of the
(magnitude-complete) COSMOS catalogue (e.g. Wallin et al. 1994) as far
as position is concerned, so that there is no magnitude-dependent bias
in the selection of the galaxies in the ENACS spectroscopic sample.
It is important to note, however, that we do not estimate the {\em
total} cluster luminosity-density, since this would require the
determination of the cluster luminosity function and its extrapolation
beyond the observed magnitude range. 

In the Jeans equation (eq.~\ref{e-jeans}) knowledge of the logarithmic
derivative $d \ln \nu / d \ln r$ is needed. We determine it
by fitting straight lines to the $\ln \nu$ vs.  $\ln r$ data-points
within intervals of $0.1 \, r_{200}$ width, spaced by $0.1 \,
r_{200}$. The procedure is repeated for each of the 64 bootstrap
estimates of $\nu(r)$, thus allowing us to estimate the r.m.s. of the
values of the fitted line slopes, and hence the confidence levels on
$d \ln \nu / d \ln r$. We checked for consistency that this
estimate of $d \ln \nu / d \ln r$ is close to the direct
(smoothed) numerical differentiation of $\ln \nu(r)$.

\subsection{Velocity dispersion profiles}
\label{a-vprof}

In order to determine the function ${\rm{<} v_r^2 \rm{>}}(r)$ for a
galaxy class with zero velocity-anisotropy, we need to deproject the
observed line-of-sight velocity dispersion profile $\sigma_p(R)$, via
the Abel equation (see, e.g., BT),
\begin{equation}
{\rm{<} v_r^2 \rm{>}}(r) = - \frac{1}{\pi \nu(r)} \int_{r}^{\infty}
\frac{d [I(R) \sigma_p^2(R)]}{dR} \frac{dR}{\sqrt{R^2-r^2}},
\label{e-abel2}
\end{equation}
which requires a smooth estimate of the product $I(R)
\sigma_p^2(R)$. We determine $\sigma_p(R)$ using the
LOWESS technique, with a smoothing length chosen to ensure consistency
with the binned estimate of $\sigma_p(R)$. Here, galaxies beyond
$r_{200}$ are not used, since some of them could be out of
equilibrium, which could bias the estimate of the cluster velocity
dispersion. The $\sigma_p$-profile is shown in Fig.~\ref{f-vdp}. We
show the best LOWESS estimate (heavy line), within the 68\% confidence
interval (dashed lines; determined with 1000 Montecarlo simulations in
the LOWESS program, see Gebhardt et al. 1994), as well as the binned
estimates.

After extrapolating the observed $\sigma_p(R)$ to $R_t$, we deproject
$I(R) \sigma_p^2(R)$ to obtain ${\rm{<} v_r^2 \rm{>}}(r)$ through
eq.~\ref{e-abel2}. The inversion is also done for 64 of the Montecarlo
simulations of $\sigma_p(R)$, in order to obtain the confidence
intervals in ${\rm{<} v_r^2 \rm{>}}(r)$ (shown in Fig.~\ref{f-vdp}). We
did not include the uncertainties in $I(R)$, since the uncertainties
in $I(R) \sigma_p^2(R)$ are dominated by the uncertainties in
$\sigma_p^2(R)$.

In the determination of $d \ln {\rm{<} v_r^2 \rm{>}} / d \ln
r$ and its confidence levels, we followed the same procedure that we
used for $d \ln \nu / d \ln r$ (see Appendix~\ref{a-dprof}).
The confidence levels are obtained by running the same procedure on
the 64 ${\rm{<} v_r^2 \rm{>}}$ profiles obtained from the Montecarlo
simulations. We check for consistency that the $d \ln
{\rm{<} v_r^2 \rm{>}} / d \ln r$ obtained by this procedure is close to
the smoothed estimate of the numerical differentiation of 
${\rm{<} v_r^2 \rm{>}}(r)$.

>From the $d \ln \nu / d \ln r$, ${\rm{<} v_r^2 \rm{>}}$, and
$d \ln {\rm{<} v_r^2 \rm{>}} / d \ln r$ profiles, we finally
determine the mass profile via the Jeans equation (eq.~\ref{e-jeans}).
The c.l. on the mass profile immediately follow from the c.l. of the
individual terms of the equation, under the simplified assumption of
mutually independent errors.  Such an assumption is certainly too
simplistic, but all we want to achieve is a rough estimate of the mass
profile c.l., to be used for illustration purposes only.

\section{The ensemble cluster, and its implied properties}
\label{a-assum}

When combining several clusters together, it is important that the
centers of all individual clusters are determined with sufficient
accuracy and in a uniform manner. We followed the procedure described
in paper III of this series, i.e. we used in decreasing order
of preference the X-ray center, the position of the brightest cluster
member near the center, the peak in the galaxy distribution, or the
biweight average of the positions of all galaxies. The positions of
the adopted centers are given in Table~A.1 in paper XI.

In our analysis, we assume that the ensemble cluster is not
rotating. Such an assumption is probably valid for individual
clusters. Rotational motions are expected (and observed) in the
collisional component of clusters (the intra-cluster gas) as a
consequence of cluster mergers (Dupke \& Bregman 2001), but not in the
collisionless components (galaxies and dark matter; see Roettiger \&
Flores 2000). Velocity gradients have been reported in clusters but
they have not been interpreted as a signature for rotation
(e.g. Biviano et al. 1996; den Hartog \& Katgert 1996).

Another assumption we make is that of steady state. This is likely to
be a valid assumption for {\em nearby} clusters. In fact, according to
Ellingson et al. (2001) the infall rate of field galaxies in clusters
at recent epochs is negligible (probably less than 1\% in the last 2
Gyr). Moreover, Girardi \& Mezzetti (2001) found no indication of
evolution in the density and velocity dispersion profiles of galaxy
clusters between $z \sim 0$ and $z \sim 0.3$. Finally, in the
currently favoured low-$\Omega_m$ cosmology, merging and matter
accretion from surrounding filaments is expected to have stopped long
ago, leaving the time for clusters to dynamically relax (e.g. Plionis
2002).

We also assume spherical symmetry. While individual clusters depart
from spherical symmetry, the 'ensemble' cluster is spherically
symmetric by construction. An analysis of cluster ellipticities is
beyond the scope of this paper, but we expect them not to be very
large for the following reasons. First, our clusters are rich, and de
Theije, Katgert, \& van Kampen (1995) showed ellipticity to decrease
significantly with increasing richness, reaching a value of $\sim 0.2$
for Abell richness $\ga 80$. Second, our clusters are nearby, and both
Melott et al. (2001) and Plionis (2002) showed cluster ellipticity to
decrease with decreasing redshift, reaching $\sim 0.3$ at $z \sim
0.05$. Third, we only considered the inner, virialized parts of
clusters, where the tidal elongation and the effects of accretion from
the surrounding Large Scale Structure are less conspicuous
(e.g. Melott et al. 2001; West \& Bothun 1990).

\vfill

\end{document}